\documentclass[twocolumn,english,aps,superscriptaddress,prb,floatfix,nofootinbib]{revtex4-2}
\usepackage[latin9]{inputenc}
\usepackage{color}
\usepackage{babel}
\usepackage{amsmath}
\usepackage{braket}
\usepackage{amssymb}
\usepackage{mathtools}
\usepackage{graphicx}
\usepackage[unicode=true,pdfusetitle,bookmarks=true,bookmarksnumbered=false,bookmarksopen=false,breaklinks=true,pdfborder={0 0 0},pdfborderstyle={},backref=false,colorlinks=true]{hyperref}
\hypersetup{linkcolor=blue,citecolor=blue,urlcolor=blue}

\graphicspath{{./}{./figs/}}

\begin{document}

\title{Quantum Spin Liquids Stabilized by Disorder in Non-Kramers Pyrochlores}

\author{Marcus V. Marinho}
\email{mmarinho@usp.br}
\affiliation{Instituto de F\'isica, Universidade de S\~ao Paulo, 05315-970, S\~ao Paulo, SP, Brazil}
\author{Eric C. Andrade}
\email{eandrade@if.usp.br}
\affiliation{Instituto de F\'isica, Universidade de S\~ao Paulo, 05315-970, S\~ao Paulo, SP, Brazil}
\begin{abstract}
We investigate the emergence of quantum spin liquid phases in pyrochlore oxides with non-Kramers ions, in which structural randomness effectively acts as a transverse field, introducing quantum fluctuations on top of the spin ice manifold. This is contrary to the naive expectation that disorder favors phases with short-range entanglement by adjusting the spins with their local environment.  We study a minimal model for a disordered quantum spin ice,  the transverse-field Ising model,  using a real-space formulation of the gauge mean-field theory.  This approach allows the inclusion of non-perturbative disorder effects exactly,  and thus to assess the stability of the spin-liquid phase with respect to the disorder.  The analysis shows that the quantum spin ice remains remarkably stable with respect to disorder up to the transition to the polarized phase at high fields, indicating that it can occur in real materials.  A Griffiths region of enhanced disorder-induced fluctuations is restricted to the immediate vicinity of this transition due to the peculiar nature of the low-energy excitations of the problem. For most of the phase diagram, an average description of the disorder captures the physical behavior well, indicating that the inhomogeneous quantum spin ice behaves closely to its homogeneous counterpart.
\end{abstract}
\date{\today}
\maketitle


\section{Introduction}

Rare-earth pyrochlores are among the most compelling families of frustrated magnets. Their structure --- a lattice of corner-sharing tetrahedra --- combined with complex crystal-field effects on the rare-earth ions \cite{gardner10}, gives rise to a plethora of novel quantum states \cite{dimatteo04,zhitomirsky12, savary12b, yan17, rau19b, schafer23,an25,lozanogomez25}, particularly in materials exhibiting magnetic easy-axis anisotropy,  as in the spin-ice compounds \cite{henley10,castelnovo12,udagawa21}. In these materials, there is no long-range magnetic order, and the ``two-in--two--out'' rule on each tethrahedron governs its disordered ground state. This local constraint is elegantly described by an emergent gauge field analogous to Gauss' law, with its magnetic correlations revealing characteristic ``pinch point'' singularities, as captured by neutron scattering experiments \cite{castelnovo08,jaulbert09, fennell09,morris09,benton16}.  In the presence of quantum fluctuations, this classical state is promoted to a Quantum Spin Ice (QSI), where the low-energy dynamics are governed by compact quantum electrodynamics, whose excitations include a gapless emergent photon, alongside massive visons and spinon excitations \cite{hermele04b,ross11,shannon12,benton12,sibille17,bin25}.

While it is not straightforward to encounter spin-ice materials with appreciable quantum fluctuation, Ref. \citep{savary17a} proposed an ingenious way to stabilize a QSI via randomness in non-Kramers doublet rare-earth ions, including the promising QSI candidates Pr$_2$Zr$_2$O$_7$, \cite{wen17,martin17,hicken25} and Pr$_2$Sn$_2$O$_7$ \cite{luo25}. In these systems, the ground-state doublet degeneracy is protected by the $D_{3d}$ point group symmetry, and it can be described by a pseudospin $1/2$ operator $\boldsymbol{S}_i$. Its component $S_i^z$ is time reversal odd, while the transverse components $S_i^{\pm}$ are symmetric under time reversal. The breaking of the $D_{3d}$ symmetry by quenched disorder directly acts as a local transverse field coupling linearly to the transverse components of the pseudospin, introducing quantum fluctuations in the problem, and paving the way to stabilize the QSI state either by chemical doping or by applying strain \citep{savary17a, roechner16, benton18a, emonts18}. Nonetheless, it is essential to check the stability of the QSI state with respect to the disorder itself, as it might also act to destabilize the phase, as it is observed in analogous situations of order by structural disorder \cite{henley89, maryasin14, andreanov15,sarkar17,andrade18,michel21}. 

In ordered magnets, the effects of disorder vary greatly. In unfrustrated systems, the ordered state is usually stable in the presence of defects, with disorder provoking mild modifications in the value of the order parameter and a broadening of the excitations \cite{av12b, vojta13}. The exception is the vicinity of a quantum critical point, where one might observe a Griffiths phase, characterized by a substantial enhancement of disorder fluctuations \cite{griffiths69, igloi05,thomas_jose14, thomas19}. For frustrated systems, disorder can stabilize phases without long-range order and with short-range entanglement,  such as spin glasses or random singlets \cite{fischer93,villain79,bhatt_lee82,fisher94, fisher95, saunders07, andreanov10,kimchi18,kawamura19}. 

In this work, we study a disordered transverse-field Ising model on the pyrochlore lattice as a minimal model for hosting the QSI. We include the effects of disorder as a random field and solve the model using a real-space implementation of the gauge mean-field theory (GMFT) \cite{savary12a,desrochers23} on finite clusters. We find that the QSI is robust and that an average description of the disorder is sufficient to characterize most of the QSI phase accurately. Overall, the effects of disorder are milder than naively suggested by previous studies of the transverse-field Ising model without frustration \cite{kovacs11}. In particular, a putative Griffiths phase exists only in the immediate vicinity of the transition between the QSI and the paramagnetic phase. We link this observation to the peculiar form of the spinon wavefunction in this problem.  

This paper is organized as follows. In Sec. \ref{sec:model}, we present the model and discuss its solution using GMFT directly in real space. In Sec. \ref{sec:sol} we present the solution of the model in finite clusters and construct its phase diagram using the average value of the spinon gap and its relative fluctuations as estimators for phase boundaries. Sec. \ref{sec:loc} discusses properties of the spinon wavefunction and its uncanny localization properties obtained from the level spacing analyses. The effects of further exchange terms and finite size effects are discussed in Sec. \ref{sec:further},  whereas Sec. \ref{sec:conclusions} concludes the paper with a discussion of our results.

\section{Model and mean-field theory}\label{sec:model}

As a minimal model for the QSI, we start with the nearest-neighbor spin-ice Hamiltonian supplemented by a transverse field

\begin{align}
\mathcal{H} = J\sum_{\langle i,j\rangle}S_{i}^{z}S_{j}^{z} - \sum_{i} h_{i} S_i^{x},
\label{eq:h}
\end{align}
where $J$ denotes the nearest neighbor exchange coupling between sites $i$ and $j$ in the pyrochlore lattice and $h_{i}$ is the random transverse field. The effects of further exchange terms will be discussed in Sec.~\ref{sec:further}. The pseudospins are written in the local reference frame of the pyrochlore lattice~\cite{ross11, rau19b}. For $h_i=0$, this model captures the classical spin-ice phase. In large fields, we observe a polarized phase with the spin pointing along the local field direction.  Previous investigations focused on the particular case of uniform fields, $h_i=h$. They showed that a QSI is stable for an infinitesimal $h$ and there is a discontinuous transition between the QSI and the polarized phase at a critical field $h_c$~\cite{savary17a,roechner16,emonts18}.

The pyrochlore lattice is composed of corner-sharing ``up" and ``down" tetrahedra~\cite{gardner10}.  For an arbitrary tetrahedron $t$, we define the ``two-in-two-out" rule, or the spin ice rule, through the charge operator
\begin{equation}
    Q_t
    =
    \eta_t
    \sum_{i\in t}
    S_i^z
    ,
\end{equation}
with $\eta_t = \pm 1$ for up/down tetrahedron.  In the absence of a transverse field,  this charge operator vanishes on every tetrahedron at zero temperature. For a non-zero transverse field, spin flip processes are allowed, which violate the spin ice rule. In particular, when the site-dependent random transverse fields are comparable to the nearest-neighbor coupling,  $h_i / J\sim 1$,  we need to resort to non-perturbative methods to solve the problem.  

In this work, we employ the parton representation introduced in Ref.~\cite{savary12a}:
\begin{equation}
    \label{eq:parton_rep}
    S^z_i = s^z_{ab}, \quad
    S^+_i = \Phi^\dagger_{a}s_{ab}^+\Phi_b,
\end{equation}
where $i$ is the pyrochlore site shared by the two tetrahedra $a$ and $b$, up and down, respectively.  The random transverse field becomes $h_i = h_{ab}$. We define the bosonic operators $\Phi_a$ ($\Phi_a^\dagger$) on the centers of the ``up" and ``down" tetrahedra of the pyrochlore lattice.  These operators dress the transverse components of the auxiliary spins and annihilate/create bosonic spinons,  which emerge as fractionalized excitations corresponding to local violations of the ice rules,  $Q_a\neq0$. In this representation, the sites of the original pyrochlore lattice become bonds of the dual diamond lattice.  Such operators satisfy the following commutation rules: $[\Phi_a, Q_b] = \Phi_a\delta_{ab}$ and $[\Phi_a^\dagger, Q_b] = -\Phi_a^\dagger\delta_{ab}$.   It is convenient to introduce the real and compact rotor variable $\varphi_a$, which is canonically conjugated to the charge operator $[\varphi_a, Q_b] = i\delta_{ab}$. With the help of this rotor variable, we can write $\Phi_a = e^{-i\varphi_a}$ and $\Phi^\dagger_a\Phi_a = 1$.  Using the parton representation, Eq.~\eqref{eq:parton_rep}, in the spin Hamiltonian, Eq.~\eqref{eq:h}, we obtain
\begin{equation}
  \label{eq:h_spinon_full}
    \mathcal{H}
    =
    \frac{J}{2}
    \sum_a
    Q_a^2
    -
    \frac{1}{2}
    \sum_{\braket{ab}}
    h_{ab}
    (
    \Phi^\dagger_a\Phi_b s_{ab}^+
    +
    \Phi^\dagger_b\Phi_a s_{ab}^-
    )
    .
\end{equation}
Eq.~\ref{eq:h_spinon_full}  is ammenable to the application of the GMFT: $\Phi^\dagger s\Phi\to \Phi^\dagger\Phi\braket{s} + \braket{\Phi^\dagger\Phi}s -\braket{\Phi^\dagger\Phi}\braket{s}$. Then, the Hamiltonian decouples into a gauge contribution analogous to a Zeeman Hamiltonian with a random field and a quadratic spinon-hopping Hamiltonian. Contrary to the spinon Hamiltonian, which couples neighboring tethraedra, the former Hamiltonian consists of a set of decoupled bonds, which is trivially soluble,  and we thus do not consider it.  Following prescription of Ref.~\cite{savary12a},  we set $\braket{s^\pm} = 1/2$ and impose the on-site constraint $\Phi_a^\dagger\Phi_a = 1$ only on average $\braket{\Phi_a^\dagger\Phi_a} = 1$.  The mean-field Hamiltonian now reads
\begin{equation}
\begin{split}
\label{eq:mf_Hamiltonian}
    \mathcal{H}
    =
    \frac{J}{2}
    \sum_a
    Q_a^2
    &
    -
    \frac{1}{4}
    \sum_{ab}
    h_{ab}
    (
    \Phi_a^\dagger\Phi_b
    +
    \Phi_b^\dagger\Phi_a
    )
    \\
    &
    +
    \sum_a
    \lambda_a
    (
    \braket{\Phi^\dagger_a\Phi_a}
    -
    1
    )
    .
\end{split}
\end{equation}
The site-dependent Lagrange multiplier $\lambda_a$ plays the role of a mass for $\Phi_a$, and it implements the average constraint in the Hamiltonian.  The random transverse field breaks the translation invariance, so we are forced to implement this Hamiltonian in real space. Using the result from Ref.~\cite{savary12a}, we promote the spinon field to a complex quantum rotor with the squared charge operator being defined as $Q_a^2\to\Pi_a^\dagger\Pi_a$, $\Pi_a = p^x_a + ip^y_a$, and represent the spinon operator as position variables $\Phi_a = x_a + iy_a$.  A straightforward substitution of the coordinates and momenta variables into the quantum spin ice Hamiltonian maps this problem onto a quantum harmonic oscillator 
\begin{equation}
    \mathcal{H}
    =
    \sum_{a\in A, B}
    \sum_{\alpha = x, y}
    \frac{(p^\alpha_{a})^2}{2m}
    +
    \sum_{a, b\in A, B}
    \sum_{\alpha = x, y}
    \frac
    {
    \alpha_{a}
    M_{ab}
    \alpha_{b}
    }
    {
    2
    }
    ,
\end{equation}
with the matrix $M$ being defined as $M_{ab} = 2\lambda_{a}\delta_{ab} - h_{ab}/2$. In this approach, the mean-field equation becomes
\begin{equation}
    \label{eq:mean_field_eqns}
    \braket{\Phi_a^\dagger\Phi_a} = \sum_n\frac{U_{an}U_{na}^T}{\omega_n} = 1,
\end{equation}
where $U_{an}=\langle a | n \rangle$ corresponds to the $a$-th entry of the $n$-th eigenvector of the matrix $M$.  The eigenvalues of $M$ are $\omega_n^2$.  We introduce the following annihilation operator
\begin{equation}
    \xi_{n}^\alpha
    =
    \sqrt{\frac{\omega_n}{2}}
    \left(
    \alpha_n
    +
    \frac{i}{\sqrt{m}\omega_n}
    p_n^\alpha
    \right)
    ,
\end{equation}
alongside with the creation operator $(\xi_n^\alpha)^\dagger$ to obtain the fully diagonalized spinon Hamiltonian:
\begin{equation}
    \mathcal{H}
    =
    \sum_n
    \sum_{\alpha = x, y}
    \omega_n
    (\xi_n^\alpha)^\dagger
    \xi_n^\alpha
    ,
\end{equation}
whose spectrum is given by the square root of the eigenvalues of the mass matrix $M$. 

The real-space implementation of the GMFT is performed in finite clusters of the diamond lattice with linear size $L $,  with $N=2L^3$ sites,  under periodic boundary conditions. The self-consistent mean-field equations $\braket{\Phi_a^\dagger\Phi_a} = 1$ are solved iteratively, ensuring that the local constraint is satisfied at every site on average.  We initialize the computation with a homogeneous Lagrange multiplier $\lambda^{(0)}_a = \lambda^{(0)}$ and iteratively update site by site employing the gradient-descent method, $\lambda_a^{(i + 1)} = \lambda_a^{(i)} + \gamma\nabla_{\lambda_a}\mathcal{H}$, until the local constraint is satisfied within a tolerance value, $|\braket{\Phi_a^\dagger\Phi_a} - 1| < 10^{-3}$. The parameter $\gamma$ sets the learning rate of the method. For transverse fields away from the quantum critical point, $\gamma = 10^{-1}$ is sufficient to ensure convergence of the mean-field equations. As we approach criticality, the energy landscape becomes increasingly flat.  Typically, $\gamma = 10^{-2}$ is sufficient to track the shallow gradient during updates.

Disorder is introduced through random on-site fields $h_i$ drawn from a bimodal distribution
\begin{equation}
\label{eq:pdf_disorder}
    p(h_i)
    =
    \frac{1}{2}
    \delta(h_i - \overline{h} - \delta h)
    +
    \frac{1}{2}
    \delta(h_i - \overline{h} + \delta h),
\end{equation}
whose mean and standard deviation are $\mu(h_i) = \bar h$ and $\sigma(h_i) = \delta h$.  We tested different disorder distributions, e.g., uniform disorder and site dilution, and the results are qualitatively the same.  Within the GMFT, the transition between the QSI and the polarized phase is continuous and characterized by the closing of the spinon gap~\cite{savary12a,desrochers23},  at odds with the discontinuous transition found beyond mean-field in the clean limit~\cite{savary17a,roechner16,emonts18}.  On general grounds, we expect that such a discontinuous uniform transition is unstable in three dimensions above a critical value of disorder due to the Imry-Ma criterion~\cite{thomas19, imry_ma}.  Assuming the transition persists beyond this point, it will be continuous, and we expect the qualitative picture captured by GMFT to be valid for moderate-to-strong disorder.

\begin{figure}[!bt]
\includegraphics[width=1\columnwidth]{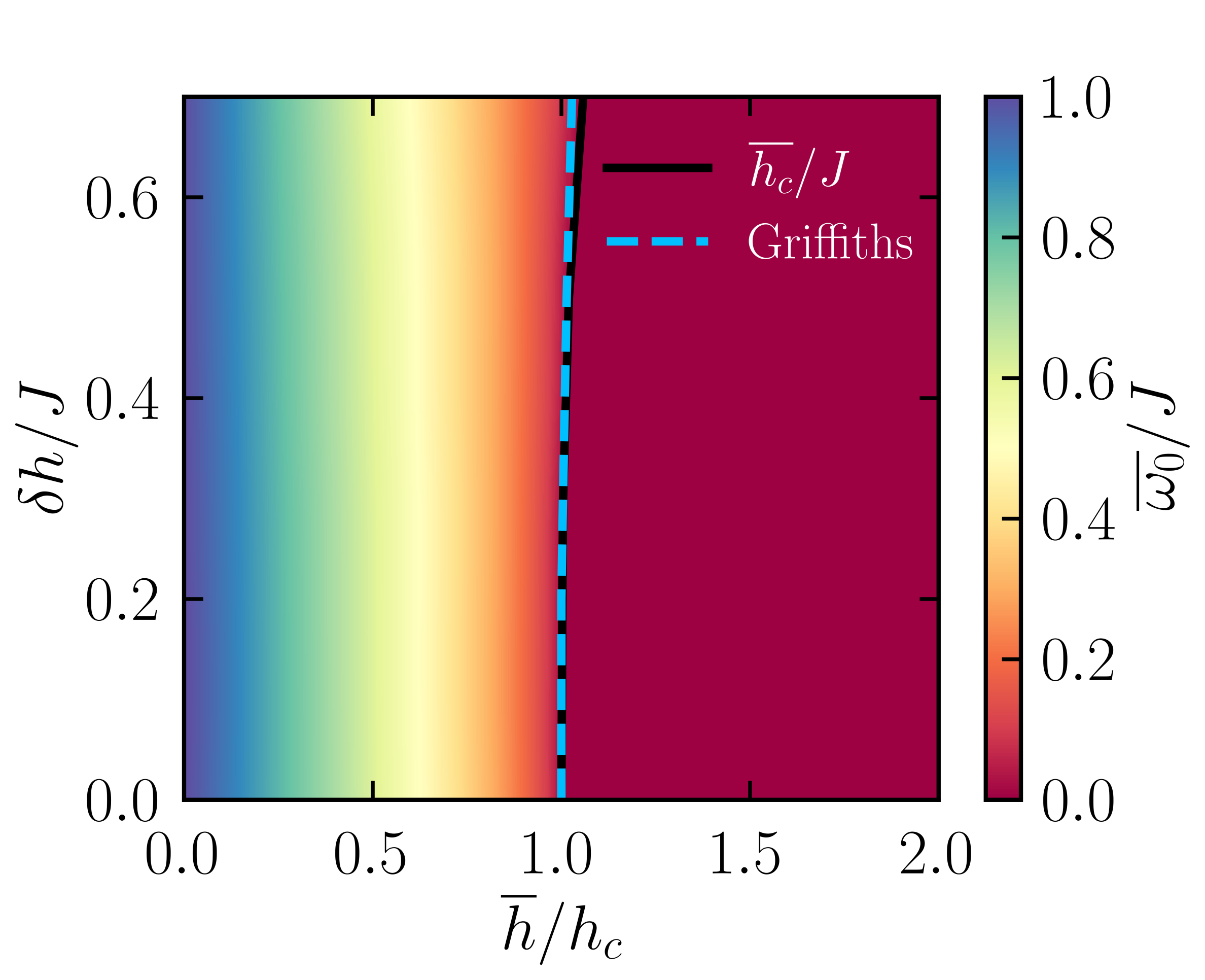}
\caption{\label{fig:pd}
Phase diagram for the transverse-field Ising model on the pyrochlore lattice. The vertical axis shows the disorder fluctuations $\delta h/J$, and the horizontal axis shows the average value of the transverse field $\overline{h}$ divided by the critical value of the field $h_c$ that separates the QSI and the polarized phase in the clean case,  $\delta h=0$.  The color code indicates the average spinon gap, which serves as an order parameter.  The values of $h_c$ are shown in Tab.~\ref{tab:dj-vs-hc}.  The results are for $L=8$. }
\end{figure}

\begin{table}[htbp]
\centering
\begin{tabular}{cc}
\hline
$\delta h/J$ & $h_c/J$ \\
\hline
0.0 & $0.711 \pm 0.001$ \\
0.1 & $0.712 \pm 0.001$ \\
0.2 & $0.713 \pm 0.001$ \\
0.3 & $0.714 \pm 0.001$ \\
0.4 & $0.722 \pm 0.001$ \\
0.5 & $0.725 \pm 0.001$ \\
0.7 & $0.752 \pm 0.002$ \\
\hline
\end{tabular}
\caption{The disorder strength $\delta h/J$ and its corresponding critical transverse field $h_c/J$ for the binomial disorder distribution in Eq.~\ref{eq:pdf_disorder}.  The results are $L=8$.}
\label{tab:dj-vs-hc}
\end{table}


\section{Phase diagram} \label{sec:sol}

\begin{figure}[!tb]
\includegraphics[width=1\columnwidth]{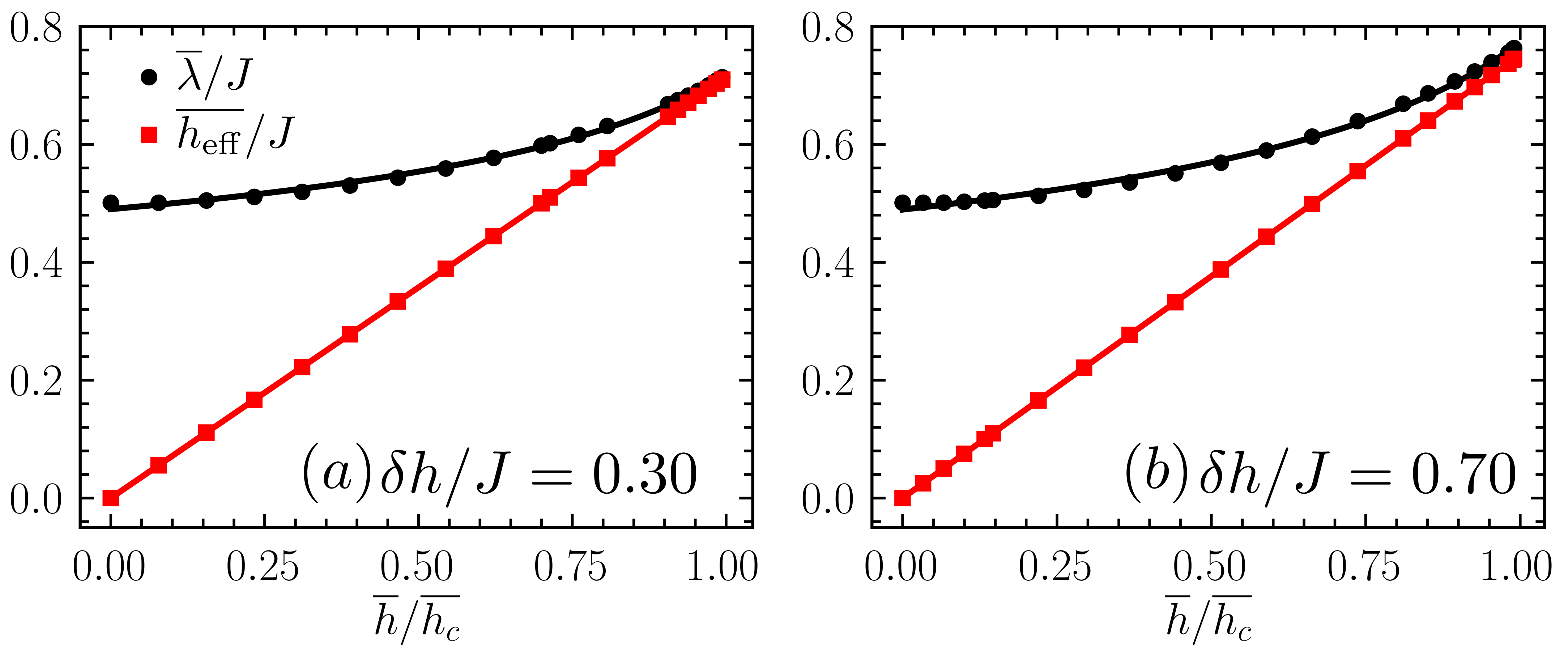}
\caption
{
Gauge mean-field results for the random transverse field Ising model of Eq.~\ref{eq:h} with a bimodal distribution of fields $h_i\in\{\overline{h}\pm\delta h\}$. The results are for a diamond lattice cluster with linear size $L=8$ ($ N=1024$ sites). The black circles show the average constraint $\overline{\lambda}/J$ and the red squares denote the average effective local field as we vary the mean transverse field normalized by the disorder-dependent critical field $\overline{h_c}$,  see Tab.~\ref{tab:dj-vs-hc}.  The solid lines are a fit to the data.  The error bars, estimated from the data variance, are smaller than the symbol sizes. 
(a) $\delta h/J = 0.30$.  Here we considered $3000$ realizations of disorder.
(b) $\delta h/J = 0.70$.  Here we considered $5000$ realizations of disorder.
}
\label{fig:mf}
\end{figure}

Within the GMFT framework,  spinons are localized in the absence of a transverse field, $h_i = 0$. As we turn on the field,  the spinons become mobile and disperse throughout the lattice. As long as the spinon gap is finite,  the QSI is stable within this formalism.  In general,  the stability of the QSI also depends on a finite vison gap ~\cite{gingras14}. From our numerics,  we extract the spinon gap as the smallest eigenvalue of the mean-field Hamiltonian in Eq.~\eqref{eq:mf_Hamiltonian},  which we dub $\omega_0$.   From the mean-field solutions in Eqs.~\eqref{eq:mean_field_eqns},  we obtain the value of the Lagrange multiplier $\lambda_i$ at each site $i$ of the diamond lattice.  As $\overline{h}$ increases, the average value $\overline{\lambda}$ also increases.  We observe that the critical point occurs at $\overline{\lambda} = \overline{h_{\rm{eff}}}$, with the effective random transverse field being defined as the sum of the fields over the sites of a given tetrahedron, $h_{\mathrm{eff}}^b = 1/4\sum_a h_{ab}$,  Fig.~\ref{fig:mf}.  In Tab.~\ref{tab:dj-vs-hc} we list the value of the critical field as a function of $\delta h$.  

We can understand this relation in the limit $\delta h \to 0$.  We focus on the spinon wavefunction,  i.e. the wavefunction corresponding to the lowest eigenvalue of the matrix $M$: $M_{ij}U_{j0}=\omega_0 U_{i0}$,  where $U_{i0}$ is the projection of the spinon wavefunction on site $i$,  see Eq.~\ref{eq:mean_field_eqns}.  Close to the transition $\omega_0 \to 0$ and we obtain $\sum_{j}M_{ij}U_{j0} \approx 0$.  For a delocalized wavefunction,  we can write $U_{j0} \sim 1/\sqrt{N}$ and $\sum_{j}M_{ij}=0$. This equation implies $\lambda_i = \sum_{j} h_{ij}/4=h_{\mathrm{eff}}^i$.  Averaging over all sites,  we obtain $\overline{\lambda} = \overline{h_{\rm{eff}}}$.  To justify the validity of this equality for finite disorder,  we only require that $U_{j0} \sim z_i$ around site $i$,  with $i$ and $j$ nearest-neighbors and $z_i$ the value of the wavefunction in the vicinity of site $i$, which we take as roughly constant.  In this fashion, our argument is also extended to localized wavefunctions.


\begin{figure}[!tb]
\includegraphics[width=1\columnwidth]{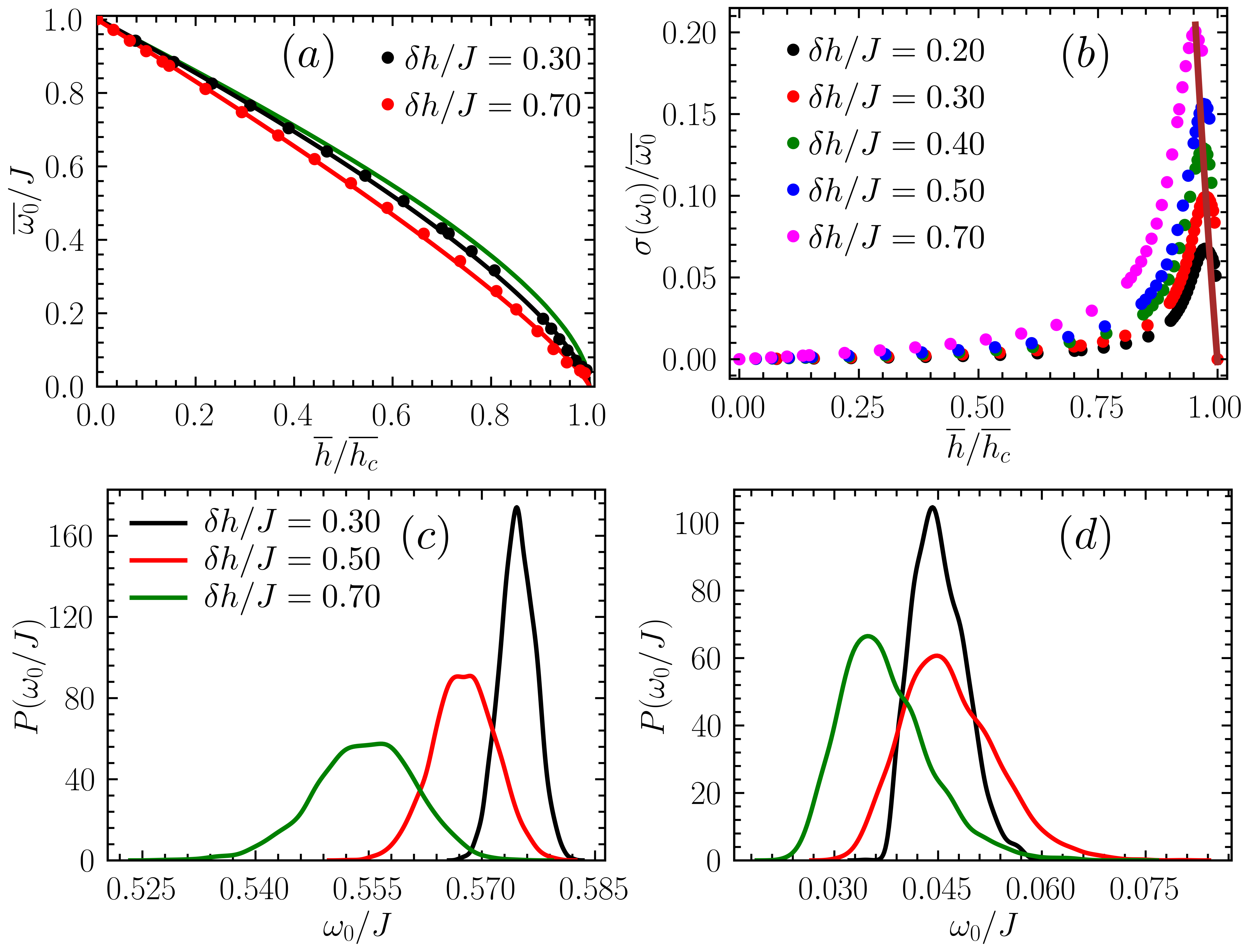}
\caption
{
Gauge mean field statistics for the spinon gap $\omega_0$ on a diamond lattice cluster of linear size $L = 8$. We used  $3000,~4000,~5000$ disorder realizations for $\delta h/J = 0.3, 0.5, 0.7$, respectively.  
(a) The average spinon gap $\overline{\omega}_0/J$ as we vary the mean transverse field normalized by the critical field $\overline{h_c}$ for $\delta h/J = 0.30$, black circles, and $\delta h/J = 0.70$, red circles.  The error bars are smaller than the symbol sizes.  The solid green line is the analytic result in the clean limit, $\omega_0/J\sim\sqrt{\lambda - h}$~\cite{savary12a}. 
(b) The spinon gap relative fluctuation for $\delta h/J = 0.20, 0.30, 0.40, 0.50, 0.70$.  The solid brown is a fit of the maximum of the curves.
(c) The spinon gap distributions at $\overline{h} = 0.50\overline{h_c}$ for $\delta h/J = 0.3, 0.5, 0.7$.
(d) The same distribution as (c), but near the quantum critical point $\overline h = 0.99\overline{h_c}$.
}
\label{fig:spinon}
\end{figure}

As previously mentioned, the transition to the polarized state in each sample can also be captured by the point at which $\omega_0 \to 0$.  Thus, an alternative definition of $h_c$ is the point where the average value of the spinon gap,  $\overline{\omega_0}$,  vanishes,  Fig.~\ref{fig:spinon}(a). In the clean case,  $\omega_0 \sim \left(h_c - h\right)^{\beta}$,  with $\beta=1/2$.  As we increase the disorder, the value of the exponent $\beta$ increases, but the spinon gap still vanishes for a finite value of the field.  The values of $h_c$ obtained within this procedure naturally coincide with the estimate from Fig.~\ref{fig:mf}.  

Because for each realization of disorder we obtain a value of $\omega_0$,  we actually construct its distribution $P\left(\omega_0 \right)$,  Figs.~\ref{fig:spinon}(c)-(d).  This quantity evolves smoothly towards the quantum critical point,  and broadens with increasing disorder strength.  Since $\omega_0$ is a positive quantity,  its distribution skews as we progressively approach the quantum critical point,  but we observe no appreciable enhancement of $P\left(\omega_0  \approx 0 \right)$,  leading to a smooth behavior of $P\left(\omega_0 \right)$, even close to $h_c$.

Close to a quantum critical point in an inhomogeneous system,  one would expect a Griffiths phase,  a region exhibiting an enhancement of the effective disorder,  which translates into power-law distributions of the local gaps~\cite{thomas19,amd09,nfl_review05,puschmann20,puschmann21,dantas22},  with important thermodynamic and spectroscopy signatures.  The physics of the Griffiths phase is usually associated with rare regions, domains in which the disorder differs significantly from its average bulk value.  Because our system sizes are modest,  it is difficult to resolve this physics,  even for a significant bare disorder $\delta h$.  To estimate the extent of a Griffiths-like region,  we then compute the relative fluctuation of $P\left(\omega_0 \right)$ as shown in Figs.~\ref{fig:spinon}(b). We set the maximum value of this quantity to the point at which the effective disorder is enhanced, linking it to a Griffiths phase. The physical motivation for this choice is that this maximum arises from samples with a spinon gap smaller than the average, indicating a weaker effective disorder in this particular realization. We see that the onset of this region of enhanced fluctuations is very close to $h_c$, and thus we have a limited window in which disorder fluctuations play a non-trivial role.

Collecting all this information from the spinon gap, we construct the phase diagram in Fig.~\ref{fig:pd}.  The phase boundary is weakly dependent on $\delta h$.  For small $\delta h$,  our results are consistent with the perturbative estimation of $h_c$ from Ref.~\cite{benton18a}.  The window of a putative Griffiths phase is also tiny, and its curvature follows that of $h_c$.  Both results indicate a weak dependence on disorder fluctuations, suggesting that an average description of the disorder is sufficient for a qualitative understanding.  In the next section, we will link this surprising finding to the peculiar nature of the spinon excitations.

\section{Spinon localization} \label{sec:loc}

\begin{figure}[!tb]
\includegraphics[width=1\columnwidth]{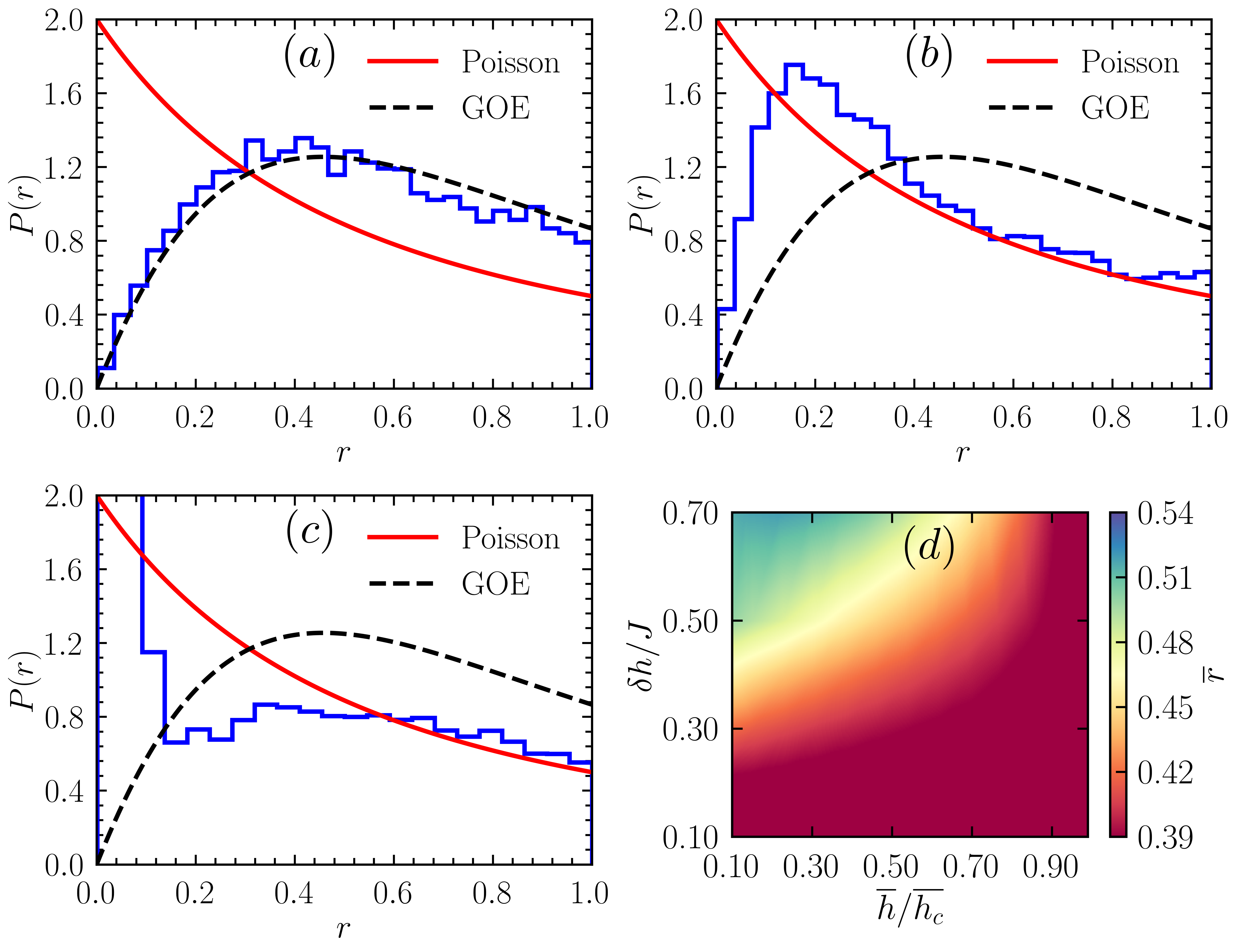}
\caption{
Results for the consecutive gap ratio statistics of the low-lying spinon modes $P(r)$.  For each disorder realization, we use the five lowest eigenenergies of Eq.~\ref{eq:mf_Hamiltonian} to compute $r$.  We consider $5000$ disorder realizations. The solid red curve is the analytical result for the Poisson regime. The dashed black curve is the analytical result for the GOE case.
(a)-(c) $P(r)$ for $\delta h/J = 0.7$ and $\overline{h}/\overline{h_c} = 0.15, 0.80, 0.99$.
(d)
Heat map of the average consecutive gap ratio for the low-lying energy levels,  $\overline{r}$,  as a function of $\delta h$ and $\overline{h}$.  The blue and red shades indicate GOE and Poisson-like statistics, respectively.
}
\label{fig:r-stats}
\end{figure}

Within the GMFT,  the spin ice ground state for zero field corresponds to a flat band of spinons,  as there is no effective hopping between different tetrahedra for the charge $Q_t$.  The spinons are then localized at each site of the diamond lattice.  For a finite field,  hopping is possible, and the spinons can now propagate throughout the lattice.  Therefore, we have a system in which disorder initially leads to the delocalization of excitations~\cite{goda06, chalker10}.  As disorder increases, we expect a reversal of this trend as Anderson localization sets in~\cite{vlad_book}.  More importantly, close to the critical field, previous studies show that low-energy excitations become localized~\cite{puschmann20,puschmann21,vojta13,zuninga13}. Therefore,  we expect a competing trend of disorder effects.  On one hand,  it delocalizes the spinon and stabilizes the QSI phase.  On the other hand, as we approach the transition, we expect the low-energy excitations to localize. 

Because our system sizes are modest,  we characterize the localization properties of the excitations via the spectral statistics, computing the consecutive gap ratio between adjacent energy levels~\cite{oganesyan07}
\begin{equation}
    r_n
    =
    \mathrm{min}
    \left(
    \frac{s_n}{s_{n + 1}}
    ,
    \frac{s_{n + 1}}{s_n}
    \right)
    ,
\end{equation}
where $s_n    =    \omega_{n + 1} - \omega_n$. Here, $\omega_n$ are the energy levels of the effective Hamiltonian in Eq.~\ref{eq:mf_Hamiltonian}.  In the localized regime, the statistical distribution of $r$ follows the Poisson distribution and reads $ P_{\mathrm{Po}}(r) = 2/(1 + r)^2 $,  with its mean value being $\overline{r}_{\mathrm{Po}}\approx 0.39$. In the delocalized regime, the system exhibits strong level repulsion.  For a time-reversal invariant system,  such as the random transverse field quantum spin ice, this leads to Wigner-Dyson distribution within the Gaussian Orthogonal Ensemble (GOE) class ~\cite{atas13}, $P_{\mathrm{O}}(r) = (27/4) (r + r^2)/(1 + r + r^2)^{5/2} $. Its mean value is $\overline{r}_{\mathrm{O}}\approx 0.54$.

In Fig.~\ref{fig:r-stats}, we display the spectral statistics obtained by analyzing the energy level spacings among the first five energy levels for $\delta h/J =0.7$.   In Fig.~\ref{fig:r-stats}(a), the disorder is strong enough to break the spectrum degeneracy near the classical spin ice manifold, and the system closely reproduces the GOE statistics already for $\overline{h}/\overline{h}_c=0.15$.  This phenomenon is analogous to inverse Anderson localization observed in flat band systems~\cite{goda06, chalker10}.  In Fig.~\ref{fig:r-stats}(b), as we approach criticality,  $P(r)$ highlights a more localized nature of the spinon wavefunction as it moves from GOE to Poisson statistics.  In Fig.~\ref{fig:r-stats} (c),  this trend is enhanced as we are very close to criticality.  The odd behavior at small $r$ happens because the energy level corresponding to the spinon gap becomes well separated from the next level for $h \to h_c$.  

All these trends are summarized in Fig.~\ref{fig:r-stats}(d), where we show the average value of the level spacing $\overline{r}$ as a function of both $\delta h$ and $\overline{h}$.   For small values of $\overline{h}$ and $\delta h$,  the spinon is localized as we are close to the classical spin ice manifold.  This implies $\overline{r}\approx 0.39$,  consistent with a Poisson distribution.  If we fix $\overline{h} \lesssim 0.5$ and increase $\delta h$, the wavefunction becomes delocalized as $\overline{r}\approx 0.54$,  consistent with the GOE limit.  Disorder thus effectively delocalizes the spinon and stabilizes the QSI by pushing the system away from the flat-band limit.  If we now fix $\delta h$ and vary the average field strength, the spinon again delocalizes, but this time it does not generally reach the GOE limit.  As we approach the critical point, the spinons become localized again due to the enhanced effective disorder.  

This dual nature of disorder makes the inhomogeneous QSI a peculiar problem.  In a usual random system,  the effective disorder increases close to the critical point,  leading to the localization of the low-energy excitations and the emergence of a Griffiths phase~\cite{vojta13,  zuninga13, puschmann20,  puschmann21}.  In finite-system simulations, this effect is amplified as the bare disorder increases.  In the present system,  the situation is distinct.  As we increase $\delta h$,  the spinon becomes more delocalized,  thus fighting against the occurrence of a Griffiths phase.  On general grounds, one still expects a Griffiths phase to appear near $h_c$~\cite{savary17a}.  However, a detailed characterization requires studies of much larger system sizes, as our current results suggest that this phase would be restricted to the immediate vicinity of the critical point.


\section{Extensions and limitations of the approach} \label{sec:further}

\begin{figure}[!tb]
\includegraphics[width=1\columnwidth]{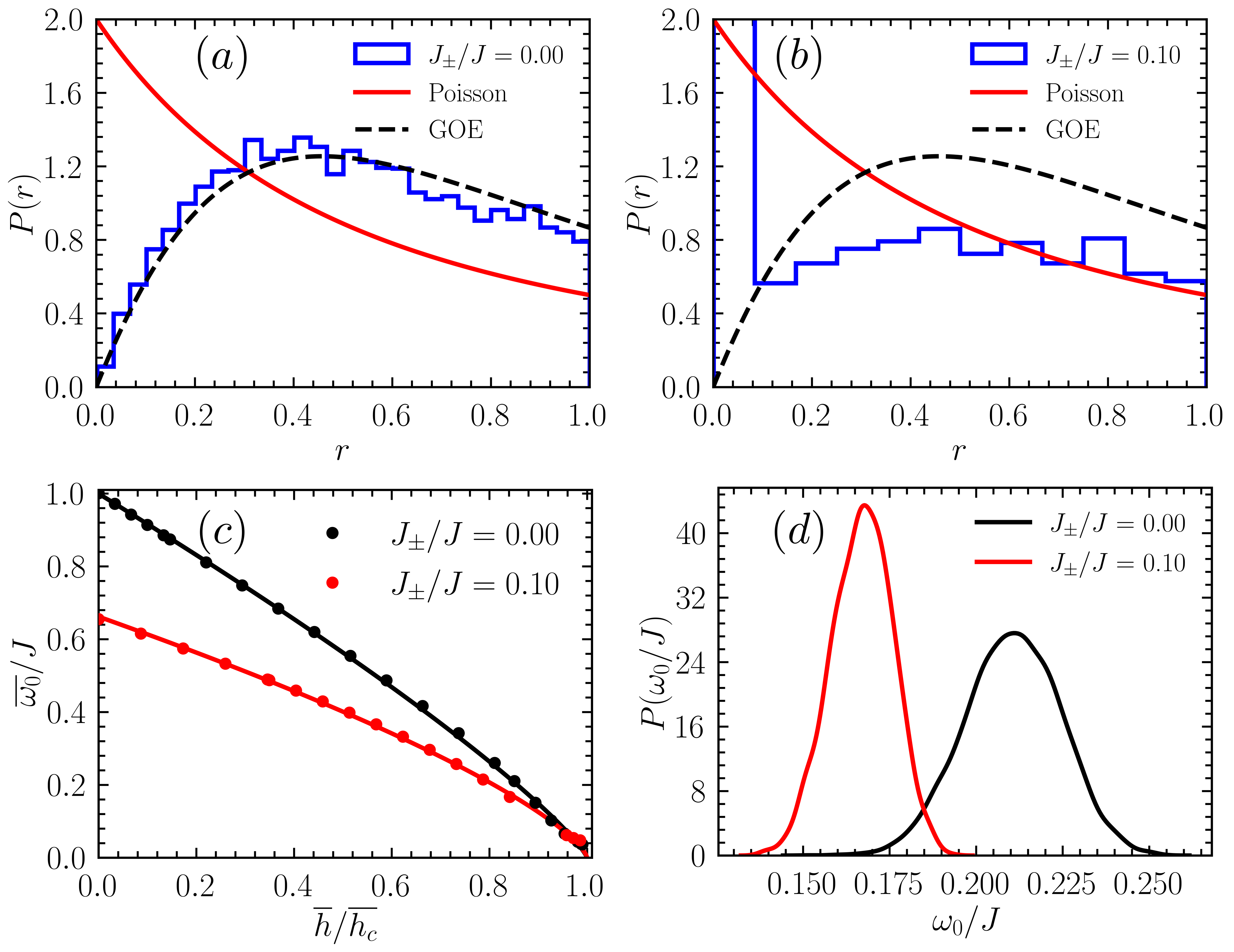}
\caption{
Effects of a finite $J_\pm$ on the gauge mean field results at disorder $\delta h/J = 0.70$ on a diamond lattice of linear size $L = 8$.
(a)-(b) The consecutive gap ratio distribution computed from the lowest five energy levels at $\overline{h} = 0.15\overline{h_c}$ for $J_\pm/J = 0.00$ (a) and $J_\pm/J = 0.10$ (b). The red solid and black dashed lines are the analytical results for the Poisson and GOE statistics, respectively.
(c) The average spinon gap as a function of the average value of the transverse field $\overline{h}$ normalized by the critical field $\overline{h_c}$. 
(d) Distribution of the spinon gap at $\overline{h} = 0.85\overline{h_c}$.  
}
\label{fig:r_distribution_Jpm}
\end{figure}

Besides the random transverse field, disordered non-Kramers pyrochlores also support additional exchange terms that couple the transverse pseudo-spin components~\cite{gingras14, rau19a, lozano_gomez24}.  It is then important to test the generality of our findings against further terms in Eq.~\ref{eq:h}.  As a minimal modification, we include the term $J_{\pm} \left( S_i^+ S_j^- + S_i^- S_j^+ \right)$, changing it to an XXZ model in a field, which might be relevant for Pr-based compounds.  For simplicity,  we consider only the average value of the coupling $J_{\pm}$.  In the GMFT language,  this term translates into a next-to-nearest neighbors hopping in the diamond lattice
\begin{equation}
    \mathcal{H}_\pm
    =
    -
    \frac{J_\pm}{4}
    \sum_{\braket{\braket{aa^\prime}} }
    \Phi_a^\dagger
    \Phi_{a^\prime}
    -
    \frac{J_\pm}{4}
    \sum_{\braket{\braket{bb^\prime}} }
    \Phi_b^\dagger
    \Phi_{b^\prime}
    .
\end{equation}
Because $J_{\pm}$ adds a hopping term to the problem, a naive guess would be that it favors the spinon delocalization.  However, its key effect is to bring the system closer to criticality, effectively renormalizing $h_c$ in a manner similar to the clean case \cite{savary12a}.   Fig.~\ref{fig:r_distribution_Jpm}(a)-(b),  shows this trend in $P(r)$.  For fixed values of the field and the disorder, $J_\pm$ changes $P(r)$ from GOE to Poisson,  similar to the trend displayed in Fig. ~\ref{fig:r-stats} as we increase $\overline{h}$ for fixed $\delta h$.   The average spinon gap in Fig.~\ref{fig:r_distribution_Jpm}(c) agrees with this picture as the gap values decrease as a function of $J_{\pm}$, indicating a destabilization of the QSI.  The distribution of the spinon energy $\omega_0$ in Fig.~\ref{fig:r_distribution_Jpm}(d) shows not only the reduction of the average value of the gap,  but also a decrease in the width of $P\left(\omega_0\right)$, signaling that it also reduces the effective disorder in the problem,  even though $\omega_0$ diminishes.  Overall,  extra exchange terms chiefly renormalize the energy scales without altering the qualitative picture emerging from Eq.  \ref{eq:h} within GMFT.  

\begin{figure}[!tb]
\includegraphics[width=1\columnwidth]{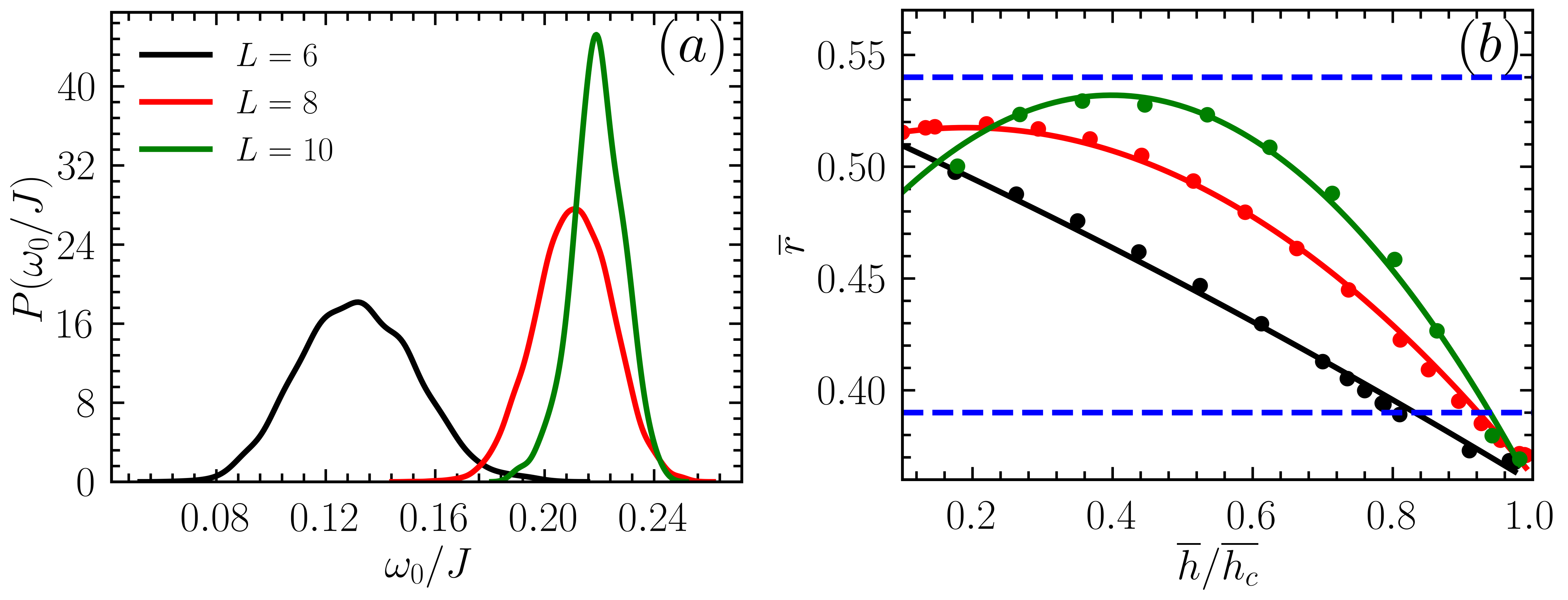}
\caption{
Finite-size effects in the gauge mean field on the diamond lattice of linear sizes $L = 6, 8, 10$, for a fixed disorder strength $\delta h/J = 0.70$. 
(a) Distribution of the spinon gap at $\overline{h} = 0.85\overline{h_c}$.  
(b) The average consecutive gap ratio $\overline{r}$ as a function of the mean transverse field normalized by the critical field. The upper and lower horizontal dashed lines mark the Poisson and GOE average values $\overline{r} = 0.54, 0.39$,  respectively. 
}
\label{fig:finite_size}
\end{figure}

As a final check of our results, we investigate finite-size effects.  The system sizes we can access with the current implementation of the real-space GFMT, based on exact diagonalization of Eq.  \ref{eq:mf_Hamiltonian}, are modest, and we only consider $L \leq 10$.  As shown in Fig.~\ref{fig:finite_size}, finite-size scaling follows the general expectations. For the distributions $P\left(\omega_0\right)$ in Fig.~\ref{fig:finite_size} (a), we observe an increase in the average spinon energy with $L$, accompanied by a decrease in the width of the distribution.  This trend is expected: as the system size increases, we reduce fluctuations, effectively moving away from the critical point and thus increasing the gap value.  The rate of change of $P\left(\omega_0\right)$ also decreases with $L$,  suggesting the convergence of the results.  In Fig.~\ref{fig:finite_size} (b), we display the average consecutive gap ratio for $\delta h/J =0.7$ as a function of $\overline{h}$.  As the system size increases,  the spinon wavefunction delocalization becomes more apparent, but no qualitative change is observed.   We thus conclude that our key observations are robust with respect to finite-size effects within the range of $L$ we can access.

To place our results in a broader perspective,  we contrast them to reference~\cite{menon20}, which studies the stability of the quantum spin-ice in the presence of disorder using exact diagonalization for small clusters.  Our mean-field approach is tailored to capture only the transition from the quantum spin liquid to the polarized phase.  Ref.~\cite{menon20} is able to probe the instability of the quantum spin-liquid towards different phases. For instance, it confirms the instability towards a spin-glass phase even at moderate disorder ~\cite{andreanov10,  benton18a}.  For stronger disorder,  Ref.~\cite{menon20} also uncovers an interesting phase consisting of random resonating-hexagons.  In general, for larger disorder, the physics becomes more local, and the authors identify the hexagon as the dominant low-energy structure of the effective model.  This is similar to a random-singlet phase found in disordered magnets~\cite{bhatt_lee82}, and its detailed characterization requires employing different approaches~\cite{sb90}. In summary,  Eq.~\ref{eq:h} describes a complex problem with competing phases, and it is not straightforward to access all of them within a single approach.

\section{Conclusions} \label{sec:conclusions}

In this work, we investigated the ground state of inhomogeneous non-Kramers pyrochlores motivated by the suggestion of Ref. \cite{savary17a} that a quantum spin ice phase is stabilized by disorder.  As a minimal model, we then considered a transverse-field Ising model with random fields.  For small fields,  this system is in the quantum spin ice phase,  whereas for large fields it is in the polarized phase.  To include the effects of disorder exactly,  we implemented a real-space version of the gauge mean-field theory and solved the problem for finite systems.  Using the average spinon gap $\overline{\omega_0}$ as an order parameter, we mapped the boundary between the quantum spin liquid and the polarized phases in the $(\overline{h},  \delta h)$ parameter space,  where $\overline{h}$ is the average value of the field and $\delta h$ its fluctuations.   We considered the relative fluctuations of $\omega_0$ as an estimator of the effective disorder in the system. We found that enhanced disorder fluctuations are restricted to the immediate vicinity of the critical point.  The full distribution $P\left(\omega_0 \right)$ is smooth for all parameters we study,  showing no signs of an accumulation of low-energy excitations in the system. Taken together, all these observations indicate that the disorder primarily renormalizes the energy scales rather than inducing strong inhomogeneous physics in the system. Since the GMFT only captures the spinon physics,  this conclusion applies only for this particular sector.

In general, we expect a Griffiths phase to emerge near the critical point in the current problem.  The limited influence of disorder fluctuations implies that the extent of the Griffiths phase in this problem is parametrically small,  as suggested in our phase diagram.  Usually,  a Griffiths phase is linked to the existence of rare regions where the local energy gap is,  for instance,  smaller than the average.  In our language, we would find puddles of almost polarized regions within the spin liquid phase.  Therefore,  the spinon wavefunction --- the ground state of our system ---  would be essentially zero in these regions,  implying it would be essentially localized in the regions favoring the spin liquid phase \cite{vojta13,  zuninga13, puschmann20,  puschmann21}.  We then conclude that this wavefunction is localized both for $\overline{h} \to 0$ and for $\overline{h} \to h_c$.  We confirm this trend numerically, investigating the degree of the localization of the spinon wavefunction using the spinon gap statistics $P(r)$.  Such a finding makes this problem special because we start with localized spinons (they are dispersionless in the spin ice phase), delocalize them by adding disorder, and ultimately localize them again as disorder increases and we reach the critical field.  We believe that this dual nature of disorder in the problem is responsible for shrinking the Griffiths phase.  We verified that our conclusions are robust to additional exchange terms and finite-size effects.  Our simple scenario suggests that the results for the clean case,  or weak disorder  \cite{savary17a,roechner16,emonts18, benton18a},  are probably sufficient to describe the quantum spin-ice phase qualitatively.

In non-Kramers systems, the spinons of the quantum spin ice involve $S^\pm$ quadrupole moments, which do not couple directly to neutron magnetic dipole moments. In contrast, emergent photons can interact inelastically with neutrons, appearing as a linearly dispersing mode that vanishes as the momentum approaches the $\Gamma$ point. Thus, we expect the dominant signal in neutron scattering experiments to be from the emerging photons \cite{chen17,  kwasigroch20,  desrochers24,  an25}. Apart from appearing in spectroscopic probes, the photons also give a $T^3$ term to the specific heat.  Because collective excitations are usually delocalized in disordered systems \cite{monthus10},  apart from the immediate vicinity of a quantum critical point,  one expects this contribution to be observed inside all the quantum spin ice phases. 

Our work also raises several questions that require further study.  Because the Griffiths phase exists on both sides of the transition, it would be interesting to study its extent within the polarized phase using, for instance, spin-wave theory in real space \cite{zuninga13, av12b}.  The gauge mean-field theory we employ predicts a continuous transition between the quantum spin ice and the polarized phase.  Treatments including fluctuations beyond mean-field predict a discontinuous transition \cite{savary17a, roechner16, emonts18}.  Thus,  including fluctuations on top of the mean-field solution would be an interesting exercise not only to investigate its effect on the order of the transition,  but also to confirm that a continuous transition is stable if disorder exceeds a critical value.  It would also be useful to investigate the stability of competing ground states in the presence of disorder --- for instance, a random singlet phase \cite{bhatt_lee82, kawamura19,  kancko25} or a spin-glass \cite{saunders07, andreanov10,  menon20} --- especially for large values of disorder where the local constraints might dominate the physics.


\acknowledgments

We thank D. Lozano-G\'omez and N. Francini for valuable discussions.
We acknowledge support by FAPESP (Brazil),  Grants No. 2021/06629-4,  2022/15453-0,  2024/02232-0,  and 2024/13170-6,  and CNPq (Brazil),  Grant No. 153799/2024-2.  ECA was also supported by CNPq (Brazil), Grant No. 302823/2022-0.

\section*{Conflict of Interest}
All authors declare no conflict of interest.

\section*{Data Availability Statement}
The data that support the findings of this study are available from the corresponding author upon reasonable request.

\section*{Keywords}
quantum spin ice, pyrochlore lattice, quantum spin liquid, disorder, transverse field Ising model, gauge mean field  theory, Griffiths phase

\bibliography{frust}

\end{document}